\documentclass[%
 reprint,
 superscriptaddress,
 amsmath,amssymb,
 aps,https://www.overleaf.com/project/5d5ae32d4618dd37df01b4f9
 pra,
floatfix,
]{revtex4-1}
\usepackage[utf8]{inputenc}

\usepackage{amsmath}
\usepackage{lipsum}
\usepackage{upgreek}

\DeclareMathOperator{\Tr}{Tr}
\usepackage{braket}

\usepackage{graphicx}
\usepackage{dcolumn}
\usepackage{bm}
\usepackage[colorlinks,linkcolor=blue,citecolor = blue,urlcolor=blue]{hyperref}
\usepackage{booktabs}
\usepackage{float}
\usepackage{csquotes}
\usepackage{abraces}

\usepackage{commath}
\usepackage{ulem}
\usepackage{braket}
\usepackage{color}
\usepackage{cleveref}
\usepackage{bbold}
\usepackage{booktabs}

\graphicspath{ {./} }
\begin{document}

\title{Spectral Properties of Transverse Laguerre-Gauss Modes in Parametric Down-Conversion}

\author{Carlos Sevilla-Gutiérrez}
 \email{carlos.sevilla@iof.fraunhofer.de}
 \affiliation{Fraunhofer Institute for Applied Optics and Precision Engineering IOF, Albert-Einstein-Strasse 7, 07745 Jena, Germany}
 \author{Varun Raj Kaipalath}
 \email{varun.raj.kaipalath@iof.fraunhofer.de}
 \affiliation{Fraunhofer Institute for Applied Optics and Precision Engineering IOF, Albert-Einstein-Strasse 7, 07745 Jena, Germany}

\author{Baghdasar Baghdasaryan}
 \affiliation{Theoretisch-Physikalisches Institut, Friedrich-Schiller-Universit\"at Jena, 07743 Jena, Germany.}
 \affiliation{Helmholtz-Institut Jena, 07743 Jena, Germany.}

\author{Markus Gr\"afe}
 \affiliation{Fraunhofer Institute for Applied Optics and Precision Engineering IOF, Albert-Einstein-Strasse 7, 07745 Jena, Germany}
 \affiliation{Institute of Applied Physics, Technical University of Darmstadt, Schloßgartenstr. 7, 64289 Darmstadt, Germany}
 
\author{Stephan Fritzsche}
 \affiliation{Theoretisch-Physikalisches Institut, Friedrich-Schiller-Universit\"at Jena, 07743 Jena, Germany.}
 \affiliation{Helmholtz-Institut Jena, 07743 Jena, Germany.}
  \affiliation{Abbe Center of Photonics, Friedrich-Schiller-University Jena, Albert-Einstein-Str. 6, 07745 Jena, Germany}
 
\author{Fabian Steinlechner}
 \affiliation{Fraunhofer Institute for Applied Optics and Precision Engineering IOF, Albert-Einstein-Strasse 7, 07745 Jena, Germany}
 \affiliation{Abbe Center of Photonics, Friedrich-Schiller-University Jena, Albert-Einstein-Str. 6, 07745 Jena, Germany}
  \email{fabian.steinlechner@iof.fraunhofer.de}
\date{\today}

\begin{abstract}
The first  color photos of the parametric down-conversion (PDC) emission cone illustrate the correlation of longitudinal- and transverse momentum in the process, i.e., wavelength-dependent emission angle of PDC photons. However, current experiments and applications are more conveniently described in terms of discrete mode sets, with the most suitable choice depending on the propagation symmetries of the experimental setting. Remarkably, despite the fact that experiments with PDC sources are becoming ever more demanding, e.g. in terms of brightness or state fidelity, a description of spectral-spatial coupling in parametric downconversion for the case of discrete modal decompositions remains elusive. 
We present a comprehensive study, in theory and experiment, of the spectral dependence of the transverse Laguerre-Gauss modes in parametric downconversion. 
Moreover, we show how the spectral and spatial coupling can be harnessed to tune the purity of the well-known orbital angular momentum entanglement. This work has implications for efficient collection of entangled photons in a transverse single mode, quantum imaging, and engineering pure states for high-dimensional quantum information processing. 


\end{abstract}


\maketitle

\section{Introduction}\label{introduction}

Spontaneous parametric down-conversion (SPDC) is a versatile resource that can be tailored to the most challenging quantum information processing tasks. The maturity of its understanding, together with the continuous evolution of photonic technology in various degrees of freedom, has provided a myriad of ways of characterizing and engineering photon sources~\cite{Anwar2021,Chen2021,Kulkarni2022,Kulkarni2018,Rozenberg_22}. They are exploited for a variety of experiments and proof-of-concept applications, including a recent demonstration of quantum computational supremacy using photons~\cite{Zhong2020}. 

One of the most attractive features of SPDC is that it quite naturally yields high-dimensional entanglement in multiple degrees of freedom. The entanglement generated via SPDC can be distributed over many modes, such that characterization of this entanglement becomes a key challenge. Studies have been carried out to quantify the large amount of entanglement in SPDC in space, time, and photon number~\cite{Roux2020,Gatti2012}. However, these rely on the Schmidt number as a figure of merit, which does not say much about the Schmidt basis, in general, apart from its dimensionality.

Studies of the spatio-temporal correlations in SPDC typically use a continuous variable description of the spatial and temporal domain, relating the emission angle and frequency of the photons created ~\cite{X_entanglement_Gatti,V_entanglement_Gatti,Horoshko2012}. Although this is more intuitive, given that the fundamental rules of energy and momentum conservation can be directly observed, most calculations result in a highly complex six-dimensional problem. 

Modal decomposition is a practical and insightful representation. It allows us to discretize a continuous variable space, and a proper basis choice can reduce the number of dimensions needed without loss of information about the state. It also acts as an ideal way to connect theory and experiment, since there exist accurate experimental techniques that can generate, manipulate, and detect such modes in space and time~\cite{Bolduc2013,Eckstein2011}. Modal decomposition has been applied to study SPDC, although most previous research has considered the spectral and spatial domains independently. These studies have addressed the regime in which these correlations are purposely minimized - either through the use of unrealistic narrowband spectral filters~\cite{Walborn_2010,Palacios2011}, loosely focused pump and collection beams in short nonlinear crystals \cite{PhysRevA.103.063508,Miatto2011, Spat_Schmidt_Miatto,Bound&Opt_OAM_Miatto} or spatially monomodal waveguides~\cite{Ansari2018}. This approach neglects the well-known correlation of the photons' spatial and spectral properties, which is markedly reflected in the famous SPDC emission rainbow~\cite{Sergienko2003}. However, factorability of the spectral and spatial domains is not always possible, as shown by Osorio \textit{et al.}~\cite{Osorio2008}. This can be detrimental when the goal is to engineer high-dimensional spatially entangled photon sources, since any distinguishability in any degree of freedom that is not considered will reduce the purity of the entangled state. Interestingly, this coupling of the spatiotemporal domain can also be used as a resource for source engineering. Torres \textit{et al.} showed how to modify spectral properties (joint spectral amplitude) through what they call spatial-to-spectral mapping~\cite{Shi2008,Valencia2007,Spat_spec_map,Schneeloch_2016} in a non-collinear phase matching configuration by modifying the transverse spatial profile of the pump beam. 

Despite the interest and relevance of efficient SPDC sources, only a handful of studies have considered the impact of spectral-spatial coupling on the discrete mode decomposition. For example, the spatio-spectral coupling is implicit to Ref. ~\cite{Bennink2010} discussion of efficient fiber coupling of photon pairs in SPDC. Further, the effect that changing the focal position of a Gaussian beam has on the temporal properties of photons was discussed in Refs. 
~\cite{Where_Molina,Bandwidth_control_Molina,Measurement_Shape_Molina}.
Despite these efforts, and to the best of our knowledge, the question of the spectrally-dependent mode decomposition has not yet been discussed in more general terms. Despite its great practical relevance for experiments, a simple, closed, and general expression that fully describes the spatio-temporal correlations of SPDC in a discrete spatial mode basis has remained elusive. To contribute to this need, we recently derived a simple expression for the biphoton state in terms of  Laguerre-Gauss modes, which is valid in a wide range of SPDC configurations~\cite{Baghi2022}. 

This article extends upon our previous work and discusses, from theory to experiment, how the spatio-spectral coupling in SPDC can be harnessed to engineer spatially entangled biphoton states on the orbital angular momentum (OAM) basis. We discuss the efficiency, purity, and fidelity of the generated states and study the spectral dependence of the OAM biphoton states. We find a simple method to achieve a bright flat spiral spectrum in a desired subspace (here demonstrated for OAM values up to $|\ell|=4$) and show that the purity of entanglement between different OAM modes can be controlled by purposely adding (suppressing) spectral distinguishability.

%

Unlike other experimental works where joint detection of photon pairs in arbitrary spatial bases are performed~\cite{Salakhutdinov_2012,DErrico2021,Valencia2020,Valencia2021}, we take advantage of advances in wavefront modulation schemes and utilize the multiplane light conversion technique (MPLC)~\cite{Morizur2010,Fontaine2019} for spatial mode analysis. 
MPLC benefits from high efficiency and mode-independent loss, thus eliminating the need for post-processing, contrary to other highly used techniques such as phase and intensity flattening~\cite{Qassim2014,Bouchard2018}. Thus, our work also reiterates the advantage of using MPLC in quantum optical experiments~\cite{Brandt2020,Markus2021,Lib2021}. 

The paper is organized as follows: In Section~\ref{theory} we review the theory of collinear SPDC and describe the spectral-spatial coupling in terms of a frequency-dependent spatial modal decomposition. In Section~\ref{experiment1} we experimentally validate our expression by means of spatial and spectral decomposition through projective measurements. Finally, in section~\ref{experiment2} we use our expression to study the spectral coupling in the OAM basis.

\section{Theory}\label{theory}

The Laguerre-Gauss (LG) mode basis is a practical choice in experimental systems that involve cylindrical symmetry. They are also a good initial approximation of the spatial Schmidt basis of the biphoton state, given that the OAM is conserved in SPDC~\cite{Mair2001}, and that the LG modes carry a well-defined OAM value. In this case the SPDC state can be expressed {as}:  
\begin{align}
 \label{eq:LG_basis_freq}
\ket{\Psi}=  \int\int d\omega_\mathrm{s}d\omega_\mathbf{i} 
 \sum_{p_\mathrm{s},p_\mathrm{i}=0}^{\infty} \sum^{\infty}_{\ell_\mathrm{s},\ell_\mathrm{i}=-\infty} C_{p_\mathrm{s},p_\mathrm{i}}^{\ell_\mathrm{s},\ell_\mathrm{i}}(\omega_\mathrm{s},\omega_\mathrm{i})&\\ \nonumber \times\ket{p_\mathrm{s},\ell_\mathrm{s},\omega_\mathrm{s}}\ket{p_\mathrm{i},\ell_\mathrm{i},\omega_\mathrm{i}},
\end{align}
where $C_{p_\mathrm{s},p_\mathrm{i}}^{\ell_\mathrm{s},\ell_\mathrm{i}}(\omega_\mathrm{s},\omega_\mathrm{i})$ denotes the mode amplitude for LG modes $\ket{p,\ell,\omega}=\int  d\bm{q}\, \mathrm{LG}_{p}^{\ell}(\bm{q})\,  \hat{a}^{\dagger}(\bm{q},\omega) \ket{vac} $ with azimuthal (or OAM) and radial indices $\ell$ and $p$, respectively. The integration limits in all the equations in the manuscript are omitted for the sake of brevity but are taken over all the space. The mode amplitudes can be calculated from the overlap integral of the two-photon amplitude (TPA) $\Phi$ and the collection modes:
\begin{align}\label{coe1}
    C^{\ell_\mathrm{s},\ell_\mathrm{i}}_{p_\mathrm{s},p_\mathrm{i}} 
    =    \iint  d\bm{q}_\mathrm{s} \: d\bm{q}_\mathrm{i} \: \Phi(\bm{q}_\mathrm{s},\bm{q}_\mathrm{i},\omega_\mathrm{s},\omega_\mathrm{i})\:[\mathrm{LG}_{p_\mathrm{s}}^{\ell_\mathrm{s}}(\bm{q}_\mathrm{s})]^*\nonumber&\\
        \times  [\mathrm{LG}_{p_\mathrm{i}}^{\ell_\mathrm{i}}(\bm{q}_\mathrm{i})]^*,
\end{align}
where $\bm{q}_{\mathrm{s},\mathrm{i}}$ and $\omega_{\mathrm{s},\mathrm{i}}$ are the transverse momenta and frequencies of the photons, respectively. Since the spatial and spectral components of the TPA are in general non-factorizable, $\Phi({\bf q}_\mathrm{s},{\bf q}_\mathrm{i},\omega_\mathrm{s}, \omega_\mathrm{i})\neq g({\bf q}_\mathrm{s},{\bf q}_\mathrm{i})\times f(\omega_\mathrm{s}, \omega_\mathrm{i})$, the coefficients $C^{\ell_\mathrm{i},\ell_\mathrm{s}}_{p_\mathrm{i},p_\mathrm{s}}$ are generally frequency dependent. We refer the reader to our recent work~\cite{Baghi2022} where we derived the expression $C_{p_\mathrm{s},p_\mathrm{i}}^{\ell_\mathrm{s},\ell_\mathrm{i}}(\omega_\mathrm{s},\omega_\mathrm{i})$ which can be used for an arbitrary spatio-temporal pump field. 

For the scope of this paper and our experimental constraints, we restrict ourselves to the case of a monochromatic Gaussian pump. Under this consideration and due to energy conservation $\omega_\mathrm{p}=\omega_\mathrm{s}+\omega_\mathrm{i}$ and OAM conservation $\ell_\mathrm{p}=\ell_\mathrm{s}+\ell_\mathrm{i}=0$ \cite{Mair2001,Baghi2022}, the SPDC biphoton state Eq.~\ref{eq:LG_basis_freq} can be reduced to:
\begin{equation}\label{eq:decomposition_gauss}
    \ket{\Psi}=
    \sum_{p_\mathrm{s},p_\mathrm{i}=0}^{\infty} \sum^{\infty}_{\ell=-\infty}\int d\Omega C_{p_\mathrm{s},p_\mathrm{i}}^{|\ell|}(\Omega) \ket{p_\mathrm{s},\ell,\Omega}\ket{p_\mathrm{i},-\ell,-\Omega},
\end{equation}
where $\Omega$ is the frequency deviation from the center frequency $\omega^0_{\mathrm{s},\mathrm{i}}$, $\omega_{\mathrm{s},\mathrm{i}}=\omega^0_{\mathrm{s},\mathrm{i}}\pm \Omega$ and $\ell=\ell_\mathrm{s}=-\ell_\mathrm{i}$. The detailed expression for $C_{p_\mathrm{s},p_\mathrm{i}}^{|\ell|}$ is shown in Eq.~\ref{eq:supp_Gauss_pump} in the appendix~\ref{supp_theory_1}. 

\section{ Spatial-Spectral decomposition}\label{experiment1}    

\begin{figure}[thbp]
\includegraphics[scale=0.7]{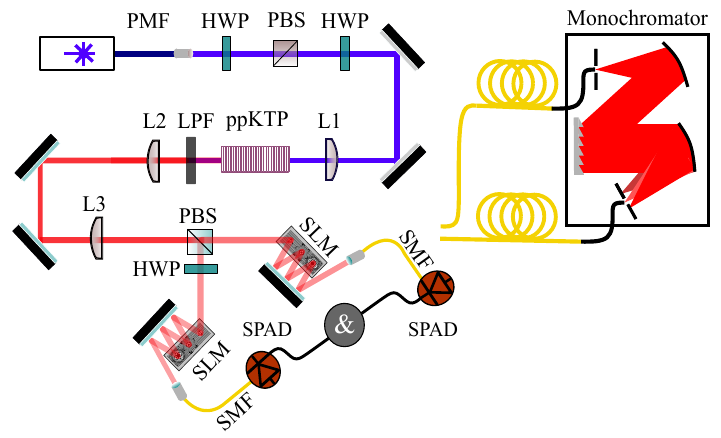}
\caption{ Experimental setup. A 405\,nm continuous-wave laser pumps a type-II ppKTP crystal. Signal and idler photons are separated by the polarizing beamsplitter (PBS). The crystal plane is imaged into the detection system consisting of the SLMs and single-mode fibers to perform joint projective measurements in the spatial domain. A monochromator is used in the path of one photon to project the state in the spectral domain. PMF: Polarization-maintaining fiber, HWP: Half-wave plate, PBS: Polarizing beam splitter, ppKTP: periodic poled potassium titanyl phosphate, LPF: Long-pass filter, SLM: Spatial light modulator, SPAD: Single-photon-counting avalanche diode, SMF: Single-mode fiber.}
\label{fig:setup}
\end{figure}

The experimental setup used to verify Eq.~\ref{eq:decomposition_gauss} is depicted in Fig.~\ref{fig:setup}. A collimated pump laser 
($\lambda_\mathrm{p}=404.8$\,nm) is focused by the lens $L_1$ ($f_1=200$\,mm) in the center of a $20$\,mm length type-II periodically poled KTiOPO4 (KTP) producing SPDC around $809.6$\,nm. The resulting pump waist is $w_\mathrm{p}\approx60$\,$\upmu$m. A long-pass filter is used to block the pump laser and the two photons are transmitted and then separated deterministically by a polarizing beam splitter (PBS) later on. The crystal plane is imaged to the input plane of the detection module by a 4-f imaging system consisting of the lenses $L_2$ and $L_3$ ($f_2=100$\,mm,$f_3=1000$\,mm). 

The spatial mode detection setup relies on MPLC to map an arbitrary spatial mode to a Gaussian mode and then is coupled to a single mode fiber (SMF), filtering the desired mode. The MPLC scheme is implemented using 3 phase modulations on liquid crystal-based spatial light modulators (SLM), and has an efficiency of approximately $40$\% for all modes considered (Appendix~\ref{section:supp_classical}). After being coupled to the SMF's, the photons are detected with single photon avalanche diodes (SPAD) and coincidence counts are recorded by a time tagger (Qtag,$1$\,ns coincidence window).  

\begin{figure}[t]
\includegraphics[scale=0.82]{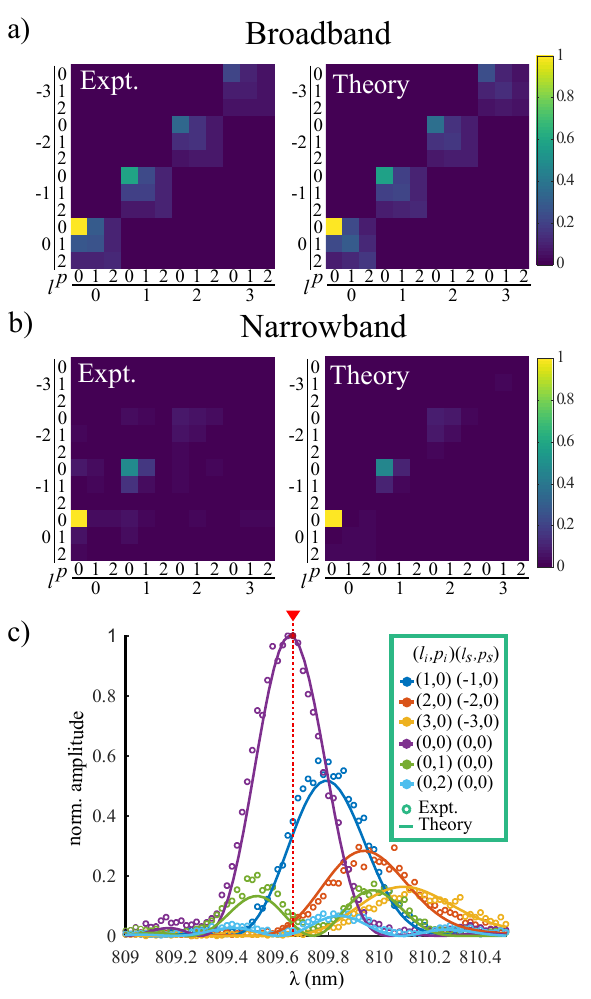}
\caption{a) Spatial mode decomposition of the spectrally broadband PDC emission in the subspace of the LG basis $p=0,1,2$ and $l=0,\pm1,\pm2,\pm3$ and for a pump waist $w_\mathrm{p}=60$\,$\upmu$m and a collection waist $w_\mathrm{s}=30$\,$\upmu$m. b) Spatial decomposition with narrowband spectral filter centered at $\lambda=809.66$ \,nm. When comparing both, we see the effect of spectral filtering in the spatial domain.  c) Joint spectrum of different joint spatial modes. We see that the center wavelength of higher-order modes is shifted from the collinear phase-matched wavelength as a result of the higher transverse momentum contributions.(Colors in online version).}
\label{fig:spat_decomp}
\end{figure}

A spatial decomposition is performed by projecting the two-photon states in all different combinations of joint measurements $\ket{\ell_\mathrm{i},p_\mathrm{i},\ell_\mathrm{s},p_\mathrm{s}}\langle \ell_\mathrm{i},p_\mathrm{i},\ell_\mathrm{s},p_\mathrm{s}|$. For this, the subspace of the LG basis with $p=0,1,2$  and $\ell=0,\pm1,\pm2,\pm3$ and a collection waist $w_\mathrm{s}=30$\,$\upmu$m was considered for signal and idler photon. 
To analyze the spectral properties, we send one of the photons to a monochromator (Andor Kymera 193 i, FWHM $\approx30$\,pm). Upon detecting the signal photon, the idler photon is automatically projected into the spectral state  $\omega_\mathrm{i}=\omega_\mathrm{p}-\omega_\mathrm{s}$ fulfilling energy conservation. 

Fig. ~\ref{fig:spat_decomp}.a. shows the experimental mode correlation matrix in the broadband regime (before connecting the monochromator). As expected, anti-correlation of the OAM and decreased spectral efficiency are observed for higher values $|\ell|$ ~\cite{Torres2003}. We repeat the same measurement, but now add a narrowband spectral filter at $\lambda_\mathrm{s}\approx809.66$\,nm, as shown in Fig. ~\ref{fig:spat_decomp}.b. The crosstalk appearing in the experimental plots between different OAM mode groups is result of misalignment due to stability issues during long time measurements. Inspection of the two figures clearly shows that spectral filtering eliminates many spatial modes generated in the down-conversion process. In fact, if we repeat the same measurement at different $\lambda_s$, the correlation matrix would be different every time (Appendix~\ref{section:supp_NB}). This can be seen directly in the measured spectrum for different spatial modes (Fig.~\ref{fig:spat_decomp}.c). 
We notice that the center wavelengths of higher-order modes tend to be shifted away from the collinear phase-matching wavelength.  This is expected, since higher-order modes carry a higher average $|\bf{q}|$, the phase mismatch is compensated at different wavelengths, similarly to temperature tuning. This dependence of the spectrum on the collected modes is due to spectral-spatial coupling. We can also see that this coupling can occur within a wavelength range where efficient spectral filtering with off-the-shelf components is not possible (in our experiment, the relevant spectral bandwidth is less than $1$\,nm). To decrease the spatiotemporal coupling, larger waist parameters are typically chosen limiting the solid angle collected from the SPDC ($|\bm{q}|\approx0$), but this results in very inefficient photon sources. 
The predictions of the theory are in good agreement with the experimental results (Figs.~\ref{fig:spat_decomp}.a-c). More realizations of the experiment were performed for different crystal lengths, pump and collection waists, and the results are displayed in appendix~\ref{section:supp_NB}. It is worth mentioning that post-processing was not necessary, since the MPLC is close to being a mode-independent technique ~\cite{Hiekkamaki2019} (at least for the chosen subspace).  

\section{Spectral dependency of the OAM states}\label{experiment2}
To see the practicality of our expression, we investigated the influence of spatial-spectral coupling on OAM anticorrelation in SPDC, which has been widely used in experiments on high-dimensional entanglement.
Restricting ourselves to the case of $p=0$, the biphoton state reads:
\begin{equation}
\ket{\Psi}=\int \mathrm{d}\Omega\sum^\infty_{\ell=-\infty} C_{|\ell|}(\Omega)\ket{\ell,\Omega}\ket{-\ell,-\Omega}.    
\end{equation}

Here we have omitted the $p=0$ for simplicity. Since the spectral amplitude $C_{|\ell|}(\Omega)$ or the collection probability $P_{|\ell|}=\int \mathrm{d}\Omega |C_{|\ell|}(\Omega)|^2$ is generally not independent of $\ell$, the amount of entanglement present in the OAM basis is reduced, compared to a maximally entangled state. To quantify the effect on spatial entanglement, we calculate the density matrix $\rho=|\Psi\rangle\langle\Psi|$ and then trace the spectral domain, $\rho_\mathrm{spatial}=\Tr_\Omega(\rho)$, which yields:

\begin{equation}
\rho_\mathrm{spatial}=\sum_{\ell,\Tilde{\ell}}A_{\ell,\Tilde{\ell}}|\ell,-\ell\rangle\langle\Tilde{\ell},-\Tilde{\ell}|.
\end{equation}

Here, $A_{\ell,\Tilde{\ell}}=\int C_\ell(\Omega)[C_{\tilde{\ell}}(\Omega)]^*d\Omega$, is the spectral overlap integral of the OAM modes $\ell$ and $\tilde{\ell}$. Evidently, poor spectra overlap reduces the magnitude of the off-diagonal elements of the density matrix, leading to loss of entanglement. It is easy to see that, the two-dimensional subspaces $|\ell,-\ell\rangle+|-\ell,\ell\rangle$ are found to be maximally entangled, since the spectral correlations $C_\ell(\Omega)$ are the same for both modes. This might not be the case for $|\Psi_{\ell}\rangle=C_{\ell}|\ell,-\ell\rangle+C_{\Tilde{\ell}}|\Tilde{\ell},-\Tilde{\ell}\rangle$ when $\ell\neq\Tilde{\ell}$.


This can be fixed by applying a quick-and-dirty approach based on the previous observations.  By simply changing the waist parameters we can directly manipulate the amount of transverse momentum of the collection modes, and with this also the spectrum. Choosing the right parameters allows us to optimize the spectral overlap $A_{\ell,\Tilde{\ell}}$, and attain a nearly flat OAM distribution $P_\ell$, thus maximizing entanglement in the relevant subspace. This approach, unlike Procrustean filtering, i.e., the intoduction of mode selective loss irrespective of spectrum~\cite{Procrustean_Zeilinger_2002}, also addresses the spectral overlap. Note that this waist choice can also be interpreted as projecting into a differently weighted distribution of $p$ states, since $LG^\ell_p(w')=\sum_p A^\ell_p LG^\ell_p(w)$, where $A^\ell_p=\int\int d\rho d\phi LG^\ell_p(w)[LG^\ell_{p'}(w')]^*$ and $w'$ and $w$ are different waist parameters ~\cite{Laguerre_waist}, thereby demonstrating the importance of considering (the often forgotten) radial mode number, when engineering spatially entangled states~\cite{Plick_radial}. This way, we are still restricted to the case $p=0$ in each OAM subspace $|\ell|$ in a redefined basis of waist $w_\ell$. 

\begin{figure}[t]
    \includegraphics[scale=0.5]{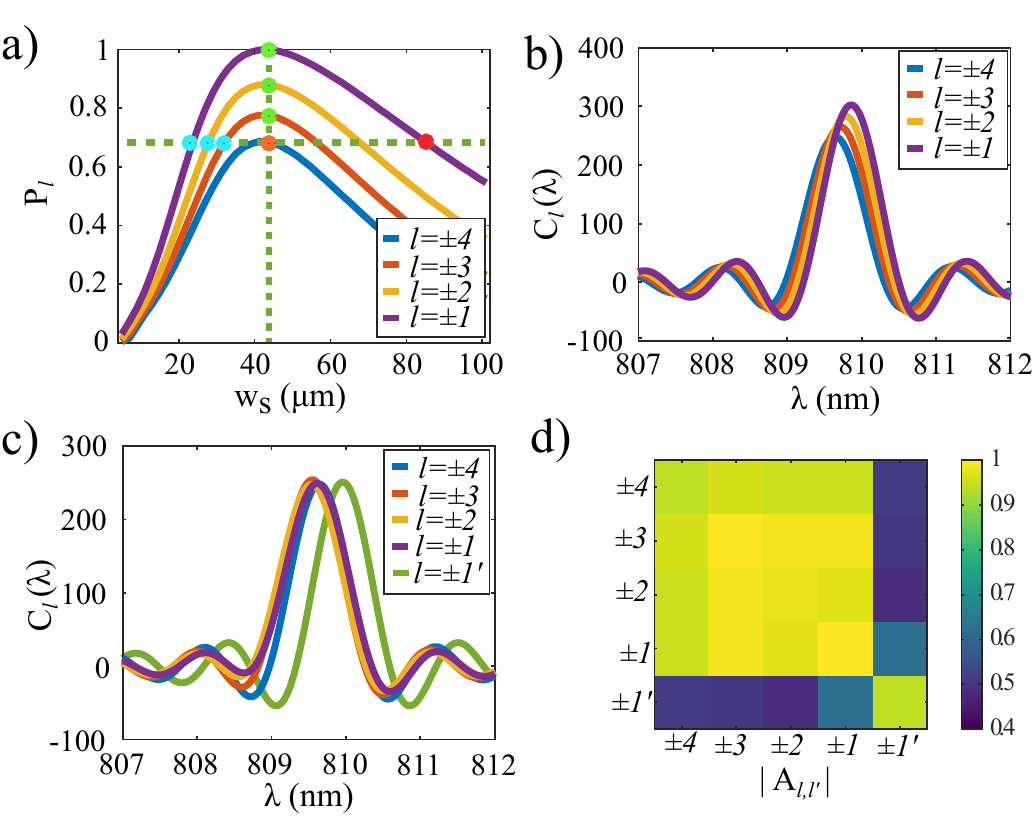}
    \caption{a) Theoretical pair collection probability  $P_{l}$ as a function of the collection waist $w_\mathrm{s}$ for the state for $|\ell,-\ell\rangle$, given the pump waist $w_\mathrm{p}=141$\,$\upmu$m. The orange dot corresponds to the optimal collection waist for the OAM mode $|l|=4$ that maximizes the collection probability. b) Theoretical spectral amplitude of the different OAM modes collected with $w_\mathrm{s}=41$\,$\upmu$m (dots on vertical dashed line in a)), showing the shift of the central wavelength and different amplitude. c) Theoretical spectral amplitude of the different OAM modes collected with different waist parameters $w_\mathrm{l}$ (dots on horizontal dashed line in a). In c) and d) $\pm1$ and $\pm1'$ represent the two different choices of collection waists $w_{1}$ and $w'_{1}$, respectively. d) Theoretical overlap of the spectra of the OAM modes. This shows how sensitive the SPDC state is with respect to the collection mode. (Colors in online version).}
    \label{fig:OAM_flatt}
\end{figure}
\begin{figure*}[t]
    \includegraphics[scale=0.55]{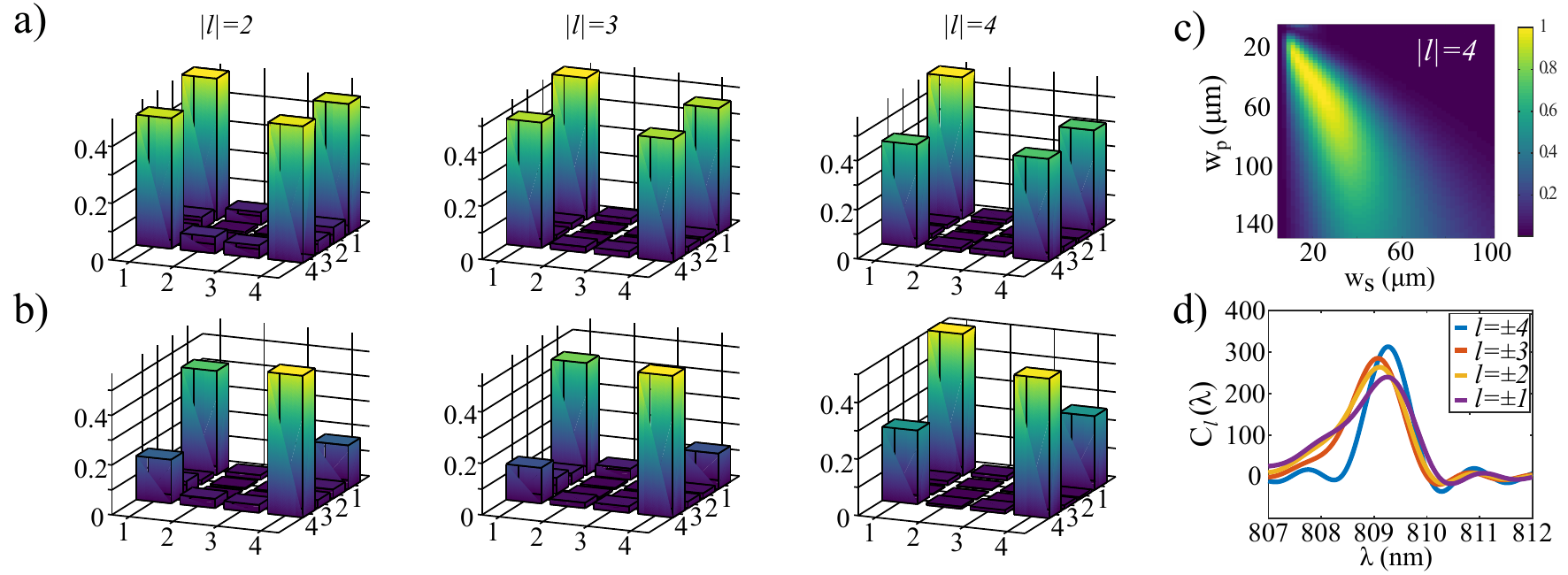}
    \caption{Absolute value of the experimentally reconstructed density matrices for the state  $\rho_{spatial}=\sum_{\ell,\Tilde{\ell}}A_{\ell,\Tilde{\ell}}|\ell,-\ell\rangle\langle\Tilde{\ell},-\Tilde{\ell}|$ for the chosen waist parameters a) $w_{1}$  and b) $w'_{1}$. It can be seen that for the latter, the off-diagonal elements have significantly reduced. c) Theoretical pair collection probability $P_{l}$ as a function of collection waist $w_s$ and pump waist $w_p$ for the state $|4,-4\rangle$. d) Theoretical spectral amplitude of the states $|\ell,-\ell\rangle$ with collection waists $w_{1}=15$\,$\upmu$m, $w_{2}=19$\,$\upmu$m, $w_{3}=21$\,$\upmu$m and $w_4=31$\,$\upmu$m. (Colors in online version). }
    \label{fig:tomo_density}
\end{figure*}

The following calculations were done considering a continuous-wave pump laser at $\lambda_\mathrm{p}=405$\,nm and a PPKTP crystal of length $L=10$\,mm. First, we select the focus waist to be used in the experiment; in this case, we set $w_\mathrm{p}=141$\,$\upmu$m. We now calculate $P_{\ell}$ for each OAM mode as a function of the collection waist $w_\mathrm{s}$, which is plotted in Fig.~\ref{fig:OAM_flatt}.a for $|\ell|=1, 2, 3$ and $4$. We then calculate the spectral amplitude of the different modes by selecting $w_\mathrm{s}$ that maximizes the mode $|\ell|=4$, $w_4=41$\,$\upmu$m (orange dot in Fig.~\ref{fig:OAM_flatt}.a). As mentioned above, if we choose the same $w_\mathrm{s}$ for all $|\ell|'s$ (green dots on the vertical dashed line in Fig.~\ref{fig:OAM_flatt}.a, the spectra are slightly shifted from each other and the amplitude also decreases for higher $\ell$, as can be seen in Fig.~\ref{fig:OAM_flatt}.b. Following the approach mentioned above, we define different collection waist parameters $w_\ell$ for each OAM mode, as shown in the red and blue dots on the horizontal dashed line in Fig.~\ref{fig:OAM_flatt}.a. The waist parameters used in the experiment are $w_{1}=25$\,$\upmu$m, $w_{2}=29$\,$\upmu$m, $w_{3}=35$\,$\upmu$m (blue dots in Fig.~\ref{fig:OAM_flatt}.a) for $|\ell|=1, 2, 3, 4$, respectively. To illustrate the consequence of spatial-spectral coupling, we also choose an erroneous waist parameter for $\ell=1$ to be $w'_{1}=85$\,$\upmu$m (we add the apostrophe to differentiate from the $w_{1}$ configuration), which has the same generation probability (red dot in Fig.~\ref{fig:OAM_flatt}.a). 
 On a more technical note, this again shows the versatility of the MPLC scheme which allows us to efficiently vary the beam waist parameter of the collection modes.  

We plot $C_\ell(\lambda)$ for these configurations in Fig.~\ref{fig:OAM_flatt}.c. As expected, the spectral overlap of the modes can be improved, by choosing a smaller waist parameters for the modes with lower OAM. In contrast, the overlap decreases when choosing the larger beam waist parameter $w'_{1}=85$\,$\upmu$m (solid green line in Fig.~\ref{fig:OAM_flatt}.c). The absolute value of the spectral overlap $|A_{l,l'}|$ is shown in Fig.~\ref{fig:OAM_flatt}.d, which has been normalized to the highest value $P_\ell$. 
Interestingly, looking at Fig.~\ref{fig:OAM_flatt}.d, we can find two configurations for $|\ell|=1$ ($w_{1}$ and $w'_{1}$) in which the collection probabilities are close to identical but one is highly entangled and the other is poorly entangled with $|\ell|=2,3$ and $4$.  
Ideally, the elements of $|A_{\ell,\ell'}|$ (the first 4 by 4 terms) should be equal to unity for all $\ell$ and $\ell'$. However, we found that this is not to be possible only by changing the waist parameters while also optimizing also for the spectral overlap. Nonetheless the approach yields a high purity and entanglement fidelity with a maximally entangled state, as will be shown in the following. It is worth mentioning that the calculation was performed with a resolution in the collection waist of $2$\,$\upmu$m, which in this tight focusing regime, can lead to non-negligible change of the amplitude and a shift of the spectrum. This also shows how sensitive the SPDC state is in the choice of the collection mode.

We tested our predictions by performing full quantum state tomography (QST) in the two-dimensional subspaces $|\Psi_{\ell,\Tilde{\ell}}\rangle=C_{\ell}|\ell,-\ell\rangle+C_{\Tilde{\ell}}|\Tilde{\ell},-\Tilde{\ell}\rangle$ for $\ell=2,3,4$ and $\tilde{\ell}=1$ for the two configurations, $w_{1}$ and $w'_{1}$. 
The chosen measurement bases are $\{|\ell\rangle ,|\Tilde{\ell}\rangle\}$, $\{|\ell\rangle \pm|\Tilde{\ell}\rangle\}$,$\{|\ell\rangle \pm i|\Tilde{\ell}\rangle\}$. 
We then reconstruct their respective density matrices $\rho_{\ell,\Tilde{\ell}}$ from the measurement results using a maximum likelihood estimation technique~\cite{Giovannini2013,James2001} and calculate the purity and fidelity of the entangled biphoton state through 
$\gamma_{\ell,\Tilde{\ell}}=\Tr\{\rho_{\ell,\Tilde{\ell}}^2\}$ and 
$F_{\ell,\Tilde{\ell}}=(\Tr\{\sqrt{\sqrt{\rho_{\ell,\Tilde{\ell}}}\rho_t\sqrt{\rho_{\ell,\Tilde{\ell}}}}\})^2$, respectively. 
Here $\rho_{\ell,\Tilde{\ell}}$ is the calculated density matrix and $\rho_t$ is the density matrix of a maximally-entangled target state
$|\Psi_{t}\rangle=|\ell,-\ell\rangle+e^{i\phi}|\Tilde{\ell},-\Tilde{\ell}\rangle$ (with normalization factors omitted for brevity).

\begin{table}[ht]
\centering 
\begin{tabular}{|c|c| c| c | c|c|} 
\hline\hline 
&  &\multicolumn{2}{c|}{Experiment} & \multicolumn{2}{c|}{Theory} \\
$\Tilde{\ell}=1$ & $\ell$ & $F_{\ell,\Tilde{\ell}}$ & $\gamma_{\ell,\Tilde{\ell}}$ & $F_{\ell,\Tilde{\ell}}$ & $\gamma_{\ell,\Tilde{\ell}}$\\ [0.5ex] 
\hline 
 &     2 & $0.953\pm0.016$ & $0.927\pm0.030$ & 0.991 & 0.983 \\ [0.5ex]  
$w_{1}$ & 3 & $0.965\pm0.012$ & $0.939\pm0.022$ & 0.992 & 0.983 \\ [0.5ex] 
 & 4 & $0.919\pm0.023$ & $0.869\pm0.040$ & 0.992 & 0.984 \\ [0.5ex]\hline 
 & 2 & $0.658\pm0.025$ & $0.564\pm0.032$ & 0.752 & 0.628 \\ [0.5ex] 
$w'_{1}$ & 3 & $0.601\pm0.024$ & $0.530\pm0.012$ & 0.755 & 0.631 \\ [0.5ex] 
 & 4 & $0.764\pm0.023$ & $0.642\pm0.024$ & 0.773 & 0.649\\ [1ex] 
\hline 
\end{tabular}
\caption{Fidelity and purity of the state $\rho_{spatial}=\sum_{\ell,\Tilde{\ell}}A_{\ell,\Tilde{\ell}}|\ell,-\ell\rangle\langle\Tilde{\ell},-\Tilde{\ell}|$. In the first column, $w_{1}$ and $w'_{1}$ represent the two choices of waist parameter for $\Tilde{\ell}=1$.}
\label{table:FP} 
\end{table}


Fig.~\ref{fig:tomo_density}.a and Fig.~\ref{fig:tomo_density}.b show the absolute values of the reconstructed density matrices for both configurations $w_{1}$ and $w'_{1}$, respectively. As expected for the latter, the off-diagonal terms are considerably smaller than the on-axis elements, representing a loss of coherence due to spatial-spectral coupling. The experimental and theoretical fidelity $F_{\ell,\Tilde{\ell}}$ and purity $\gamma_{\ell,\Tilde{\ell}}$ of the different states are shown in table ~\ref{table:FP}. The uncertainty was calculated by repeating $10000$ times the reconstruction procedure adding Poissonian noise to the recorded data. As expected from the calculated spectral overlap, we obtained higher values of $F_{\ell,\Tilde{\ell}}$ and $\gamma_{\ell,\Tilde{\ell}}$ for the $w_{1}$ configuration and significantly decreased for $w'_{1}$. The results show good agreement with our calculations. The slight discrepancy can be due to possible underestimation of the state purity and fidelity in the maximum-likelihood estimation method and/or several experimental factors such as, not fully corrected static aberrations of the SLMs, and high precision requirement when working with small beam waists. 
Although the results were not far away from the theoretical prediction even under experimental errors, our expression allows us to upper-bound the property of interest, in this case the purity and fidelity with a maximally entangled state. When the focus of the pump and collection modes is carefully chosen, it is possible to control the purity and amount of entanglement of the state. 

We also show that we can get an increase in brightness by a factor of $\approx2$ by focusing the pump beam. This is shown in Fig.~\ref{fig:tomo_density}.c, where the pair collection probability $P_4$ is displayed as a function of the pump and collections waist parameters, $w_\mathrm{s}$ and $w_\mathrm{p}$, respectively. By applying the same method as before and choosing a smaller pump waist $w_\mathrm{p}=75$\,$\upmu$m, the required collection waists also shift to smaller values ($w_{1}=15$\,$\upmu$m,$w_{2}=19$\,$\upmu$m, $w_{3}=21$\,$\upmu$m and $w_4=31$\,$\upmu$m). From the spectral amplitude of this configuration (plotted in Fig.~\ref{fig:tomo_density}.d), we can see that in the tight-focusing regime the spectrum of the photons is no longer sinc-shaped and the spectral overlap slightly decreases. Nevertheless, fidelities $F>0.97$ can still be achieved for any combination of $|\ell|=1,2,3$ and $F>0.94$ for $|\ell|=4$. This results in the limitation of this method.  

To conclude, we note that a more general version of this method can be formulated by considering the collection modes as superpositions of the radial index modes, rather than simply changing the waist parameters. The collection modes for idler and signal photons can be written as $\ket{u}=\sum_{p_\mathrm{i}} A_{p_\mathrm{i}}\ket{{p_\mathrm{i}}}$ and $\ket{v}=\sum_{p_\mathrm{s}} B_{p_\mathrm{s}}\ket{{p_\mathrm{s}}}$ ,respectively. The biphoton OAM state is $\ket{u_\ell}\ket{v_{-\ell}}=\sum_{p_\mathrm{i},p_\mathrm{s}} A_{p_\mathrm{i}}B_{p_\mathrm{s}} \ket{p_\mathrm{i},\ell} \ket{p_\mathrm{s},-\ell}$.

The spectral amplitude can be calculated through $C_{u,v}^\ell(\lambda)=\sum_{p_\mathrm{i},p_\mathrm{s}} A_{p_\mathrm{i}}B_{p_\mathrm{s}} C^\ell_{p_\mathrm{i},p_\mathrm{s}}(\lambda)$. This approach allows us to control the spectral correlations by selecting appropriate values of the coefficients $A_{p_\mathrm{i}}$ and $B_{p_\mathrm{s}}$. 
As an example, in appendix~\ref{section:supp_optimization}, we show how this can be applied to optimize brightness and spectral separability for the OAM modes $|\ell|=0,1,2$. This can be further generalized by optimizing the pump mode, which can also be achieved by using our more general expression~\cite{Baghi2022}.

\section{Discussion}

Our work shows the potential and importance of considering the spectral and spatial coupling when engineering quantum photon sources. We have experimentally verified our theoretical model and addressed the unexplored spectral dependence of the OAM basis. We showed how the spatial-spectral coupling can be used to control the purity of spatial entanglement. The method described in Section ~\ref{experiment2}, although only proved for two-dimensional subspaces, can be easily extended for high-dimensional entanglement. Furthermore, we described a more general method that can be used to shape the spectral properties through spatial filtering, which we applied to ensure spectral indistinguishability for a subset of OAM modes. All of this shows how versatile and powerful is the spatial decomposition description of SPDC for spatial and spectral control of the biphoton state.  Our work also illustrates some limitations of a description in terms of Schmidt modes. We have shown, for the LG-mode basis, that the mode-decomposition is frequency dependent, which implies that the spatial Schmidt modes will be frequency dependent. We leave the problem of determining the  full spatio-temporal Schmidt modes in the combined transverse momentum and frequency space as subject of follow-up research. 

\section*{Acknowledgements}
The presented results were partially acquired using devices funded by the German Federal Ministry of Education and Research of Germany (BMBF) through the project Quantum Photonics Labs (QPL) with the funding ID 13N15088. We acknowledge the support from the European Union's Horizon 2020 Research and Innovation Action under Grant Agreement No. 899824 (FET-OPEN and SURQUID). Funding from the Carl Zeiss Stiftung is greatly acknowledged.

\appendix
\section{Theory}\label{section:supp_theory}
The biphoton state of SPDC can be described by the expression:
\begin{align}\label{SPDC:freespace}
    \ket{\Psi} = \iint & d\bm{q}_\mathrm{s} \: d\bm{q}_\mathrm{i} \:d\omega_\mathrm{s} \: d\omega_\mathrm{i}\: \Phi(\bm{q}_\mathrm{s},\bm{q}_\mathrm{i},\omega_\mathrm{s},\omega_\mathrm{i})\nonumber\\&
  \hat{a}^{\dagger}_\mathrm{s}(\bm{q}_\mathrm{s},\omega_\mathrm{s})\:\hat{a}^{\dagger}_\mathrm{i}(\bm{q}_\mathrm{i},\omega_\mathrm{i})\ket{vac}.
\end{align}
where $\Phi({\bf q}_\mathrm{s},{\bf q}_\mathrm{i},\omega_\mathrm{s}, \omega_\mathrm{i})$ is the so-called two-photon amplitude (TPA). It encloses all the spatial-temporal correlations and is given by:
\begin{equation} \label{eq:modefunction}
\Phi({\bf q}_\mathrm{s},{\bf q}_\mathrm{i},\omega_\mathrm{s}, \omega_\mathrm{i})\sim  E_\mathrm{p}({\bf q_\mathrm{i}+q_\mathrm{s}},\omega_\mathrm{p}) \text{sinc}\left(\frac{\Delta k L}{2}\right)
\end{equation}

Here, $\Delta k= k^z_\mathrm{p}-k^z_\mathrm{s}-k^z_\mathrm{i}$ is the longitudinal phase mismatch and $E_\mathrm{p}(\bf q_\mathrm{s}+q_\mathrm{i},\omega_\mathrm{p})$ is the amplitude of the pump beam in the spectral ($\omega_\mathrm{p}$) and transverse momentum ($\bf q_\mathrm{p}=q_\mathrm{s}+q_\mathrm{i}$) space.

Following the work from ~\cite{Miatto2011} we project the spatial domain of Eq. ~\ref{SPDC:freespace} in the Laguerre-Gauss (LG) basis  ($\langle \ell_\mathrm{s},p_\mathrm{s},\ell_\mathrm{i},p_\mathrm{i}|\Psi\rangle$), which runs over the radial $p$ and azimuthal $\ell$ indices of the signal and idler photons. The LG basis is a good initial approximation of the spatial Schmidt basis of the biphoton state, given that the orbital angular momentum (OAM) is conserved in SPDC ~\cite{Mair2001}, and that LG modes carry well defined OAM (given by the index $\ell$). While considering an arbitrary pump field in its transverse profile, it can also be decomposed as a superposition of LG modes given by $E_\mathrm{p}({\bf q}_\mathrm{p})=\sum_{p_\mathrm{p},\ell_\mathrm{p}}C_{p_\mathrm{p}}^{\ell_\mathrm{p}}\mathrm{LG}_{p_\mathrm{p}}^{\ell_\mathrm{p}}$. After this projection the state can be written as:
\begin{align}\label{eq:decomposition}
    \ket{\Psi}=
    \sum_{p_\mathrm{s},p_\mathrm{i}=0}^{\infty} \sum^{\infty}_{\ell_\mathrm{p},\ell_\mathrm{s},\ell_\mathrm{i}=-\infty}\int d\Omega_\mathrm{s} d\Omega_\mathrm{i}\nonumber\\ 
    C_{p_\mathrm{p},p_\mathrm{s},p_\mathrm{i}}^{\ell_\mathrm{p},\ell_\mathrm{s},\ell_\mathrm{i}}(\Omega_\mathrm{s},\Omega_\mathrm{i}) \ket{p_\mathrm{s},\ell_\mathrm{s},\Omega_\mathrm{s}}\ket{p_\mathrm{i},\ell_\mathrm{i},\Omega_\mathrm{i}}.
\end{align}
Here, we have used the notation $\omega_j=\omega^0_{j}+\Omega_j$, where $\Omega_{j}$ is the frequency shift from the central frequency $\omega^0_{j}$. The probability amplitude $C_{p_\mathrm{p},p_\mathrm{s},p_\mathrm{i}}^{\ell_\mathrm{p},\ell_\mathrm{s},\ell_\mathrm{i}}(\Omega_\mathrm{s},\Omega_\mathrm{i})$ is the solution of the overlap integral
\begin{align}\label{eq:Slp}
C^{\ell_\mathrm{p},\ell_\mathrm{i},\ell_\mathrm{s}}_{p_\mathrm{p},p_\mathrm{i},p_\mathrm{s}}(\Omega) = N\int d{\bf q}_\mathrm{s} d{\bf
q}_\mathrm{i} \text{sinc}\left(\frac{\Delta k L}{2}\right)
\mathrm{LG}^{\ell_\mathrm{p}}_{p_\mathrm{p}}\\\nonumber 
\mathrm{LG}^{\ell_\mathrm{i}}_{p_\mathrm{i}}\left({\bf q}_\mathrm{s}\right)^{*} \mathrm{LG}^{\ell_\mathrm{s}}_{p_\mathrm{s}}\left({\bf
q}_\mathrm{i}\right)^{*},
\end{align}
which gives all the information of the spatial-spectral correlations. Here, $N$ is the normalization factor. A close expression for this coefficient has been reported under special considerations such as spectrally narrowband or thin-crystal approximation \cite{PhysRevA.103.063508} (Rayleigh range $Z_r > $ Crystal length $ L$) . In our recent work ~\cite{Baghi2022} we have derived a semi-analytical expression for $C_{p_\mathrm{p},p_\mathrm{s},p_\mathrm{i}}^{\ell_\mathrm{p},\ell_\mathrm{s},\ell_\mathrm{i}}(\Omega_\mathrm{s},\Omega_\mathrm{i})$, where only two approximations were applied: the paraxial approximation ($k\gg|q|$) and small frequency shifts with respect to the center frequency ($\Omega\ll\omega_0$). Taking this into account, the longitudinal wave vector can be expanded as follows:
\begin{equation}
k^z_\mathrm{j}=\sqrt{k^2_\mathrm{j}-|\bm{q_\mathrm{j}}|^2}\approx k_{\mathrm{j},0}+\frac{\Omega_\mathrm{j}}{u_{\mathrm{j},0}}+\frac{G_{\mathrm{j},0}\Omega_\mathrm{j}^2}{2}-\frac{|\bm{q_\mathrm{j}}|^2}{2k_{\mathrm{j},0}},
\label{eq:approx}
\end{equation}
where $k_{\mathrm{j},0}$, $u_{\mathrm{j},0}$ and $G_{\mathrm{j},0}$ are the wave vector, group velocity and group velocity dispersion, respectively, evaluated at the center frequency $\omega_{\mathrm{j},0}$ . From Eq.~\ref{eq:approx} we can see directly that an increase in the transverse momentum $|\bm{q_\mathrm{j}}|$, related to the detection spatial mode, will necessarily affect the spectral state of the collected photons such that the PM condition is satisfied.
\subsection{SPDC expression for a CW Gaussian pump beam}\label{supp_theory_1}
We consider a continuous wave Gaussian beam given by:
\begin{eqnarray*}
   \mathrm{V}(\bm{q}_\mathrm{s}+\bm{q}_\mathrm{i}) \: = \: \frac{w_p}{\sqrt{2 \pi}}\:\exp{\biggl(-\frac{w_p^2}{4}|\bm{q}_\mathrm{s}+\bm{q}_\mathrm{i}|^2\biggr)\delta(\Omega_s+\Omega_i).}
  \end{eqnarray*}

Under this consideration, we can simplify Eq.~\eqref{eq:Slp} using $\ell=\ell_\mathrm{s}=-\ell_\mathrm{i}$ and $\Omega=\Omega_\mathrm{s}=-\Omega_\mathrm{i}$, and which reads ~\cite{Baghi2022}: 
\begin{align}\label{eq:supp_Gauss_pump}
   C^{|\ell|}_{p_s,p_i}(\Omega)
     \propto & \sum_{s=0}^{p_s}\sum_{i=0}^{p_i} \frac{w_p}{\sqrt{2}}\: (T_s^{p_s,|\ell|})^* \:(T_i^{p_i,|\ell|})^* \nonumber\\
   &  
  \int_{-L/2}^{L/2}dz\:\exp{\biggl[i z\biggl(\frac{\Omega}{u_i}-\frac{\Omega}{u_s}-\frac{\Omega^2}{2}(G_i+G_s)\biggl)\biggl]}\nonumber\\
   &  \frac{D^{-\abs{\ell}}}{H^{1+s}\: B^{1+i}}\: {_2}{\Tilde{F}}_1\biggl[1+s,1+i, 1-\abs{\ell},\frac{D^2}{H \,B
  }\biggl],
\end{align} 
where the function ${_2}{\Tilde{F}}_1$ is known as the \textit{regularized} \textit{hypergeometric} function \cite{Hypergeometric2F1}. The missing coefficients are given by
\begin{eqnarray*}
 D &=& -\frac{ w_p^2}{4}-iz\frac{1}{2k_p},\\[0.1cm]  
 H &=& \frac{w_p^2}{4}+\frac{w_s^2}{4}-i z\frac{k_p-k_s}{2k_p k_s},             \\[0.1cm]
  B& = &  \frac{w_p^2}{4}+\frac{w_i^2}{4}-i z\frac{k_p-k_i}{2k_p k_i},  \\[0.1cm]
       T_u^{p,\ell} &=& \sqrt{\frac{p!\,(p+|\ell|)!}{\pi}}\,
   \biggr(\frac{ w}{\sqrt{2}}\biggl)^{2u+|\ell|+1}\,\frac{(-1)^{p+u}}{(p-u)!\,(\abs{\ell}+u)!}
\end{eqnarray*}

Here we used the fact that $C^{\ell,-\ell}_{p_s,p_i}(\Omega)= C^{-\ell,\ell}_{p_s,p_i}(\Omega)= C^{|\ell|}_{p_s,p_i}(\Omega)$. Note that for the case of type II phase matching, the state is asymmetric in the radial dimension and the $p_s$ and $p_i$ cannot be exchanged, for example $C^{|\ell|}_{0,2}(\Omega)\neq C^{|\ell|}_{2,0}(\Omega)$. Clearly this is also the case for $\ell_s$ and $\ell_i$ when $\ell_p\neq0$.   

\section{Characterization of the mode projection system}\label{section:supp_classical}
\begin{figure}[ht]
\includegraphics[ scale=0.9]{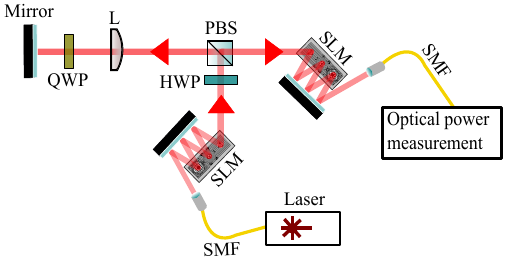}
\caption{ Characterization setup. A 808-nm continuous wave single-mode fiber coupled laser is incident on the MPLC system generating the desired spatial modes. The generated mode is send to the other MPLC sytem after the reflection from the mirror and polarization rotation achieved using the quarter wave plate. The optical power coupled to the single mode fiber after projecting to different spatial modes for each input spatial mode is measured using a photo diode and a power meter. QWP: Quarter-wave plate, HWP: Half-wave plate, PBS: Polarizing beam splitter,  SLM: Spatial light modulator, SMF: Single-mode fiber. (Colors in online version).}
\label{fig:classical setup}
\end{figure}

For the spatial mode projections, we used the Multiplane Light Conversion (MPLC) scheme. To characterize the performance of the system, we used one of the MPLC system to generate the spatial modes that are of interest in the experiment and the other MPLC to detect those modes. The experimental setup is shown in Fig.~\ref{fig:classical setup}. A collimated Gaussian beam of approximately 750 $\mu$m beam waist from a single mode fiber coupled 808 nm laser diode is incident on the MPLC system consisting of three planes of phase modulation and free space propagation, which is implemented by the spatial light modulator (SLM) and a mirror. This system generates the desired modes with a beam waist of 400 $\mu$m. The generated modes are fed back into the other MPLC system using a 4-f system, where a mirror is placed in the focal plane of the lens ($f= 1000$\,mm) and the polarization is rotated by 90$^{\circ}$ by passing through the quarter-wave plate twice, with the fast axis oriented at 45 $^{\circ}$. Thus, the beam is reflected by the PBS and the input mode is mapped to the first plane of the other MPLC with a beam waist of 400 $\mu$m. A HWP is placed in the beam path so that the polarization of the incident field is changed to the polarization state that the SLM can modulate ~\cite{C.Rosales-Guzman2017}. In the detection MPLC, the reverse transformation as the previous MPLC is performed wherein, after the three planes of phase modulation, the correct mode is transformed to fundamental Gaussian mode that gets coupled efficiently into the single-mode mode fiber. When other orthogonal modes are incident, the transformation also produces a field profile that is orthogonal to the fundamental Gaussian mode and hence does not couple to the single-mode fiber. We characterized the detection of 21 different modes of the LG basis which we used in the mode decomposition of the SPDC photons. 
\newline

\begin{figure}[ht]
\includegraphics[ scale=0.82]{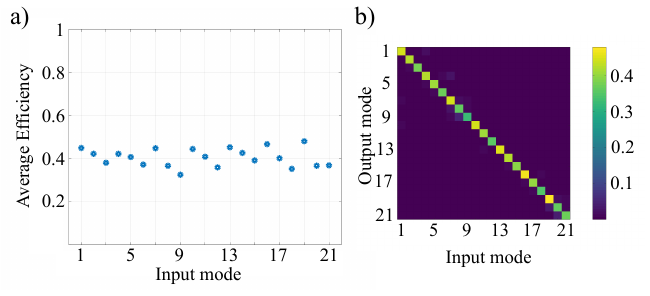}
\caption{Detection of LG modes: (a) Detection efficiency,(b) Modal crosstalk matrix, for 21 different modes of the Laguerre-Gauss basis. The mode number represents specific modes as follows: 1- LG$_0^1$, 2- LG$_1^{1}$, 3- LG$_2^1$, 4- LG$_0^{2}$, 5- LG$_1^2$, 6- LG$_2^{2}$, 7- LG$_0^{3}$, 8- LG$_1^3$, 9- LG$_2^{3}$, 10- LG$_0^0$,  11- LG$_1^{0}$, 12- LG$_2^0$, 13- LG$_0^{-1}$, 14- LG$_1^-1$, 15- LG$_2^{-1}$, 16- LG$_0^{-2}$, 17- LG$_1^{-2}$, 18- LG$_2^{-2}$, 19- LG$_0^{-3}$, 20- LG$_1^{-3}$, 21- LG$_2^{-3}$. (Colors in online version).} 
\label{fig:classical results}
\end{figure}

For a specific input mode, the power coupled into the single-mode fiber after the transformation from the detection MPLC, are measured for different detection modes. From these measurements, the modal crosstalk matrix can be created which is used to evaluate the performance of the detection system. We calculate visibility (V), which quantifies the quality of the transformations from the crosstalk matrix as follows:
\begin{equation} \label{eq:14}
V= \frac{\sum_{i}{P_{ii}}}{\sum_{ij}P_{ij}}.
\end{equation}
Here, P$_{ii}$ are the diagonal elements of the crosstalk matrix and P$_{ij}$ corresponds to all the entries in the crosstalk matrix. We also calculate the detection efficiency for each of the modes as the ratio between the power coupled to the SMF and the input power for the right combination of the input mode and the detection holograms. The modal cross-talk matrix and the plot of detection efficiency for the 21 modes are shown in Fig. ~\ref{fig:classical results}.

The clear diagonal in the crosstalk matrix shows that the detection system performs quite well. Ideally, visibility should be 1, in case of a perfect detection system. We obtain a visibility of approximately 0.91. The main reason for the reduction is visibility is from the $p$ modes as the orthogonality of the $p$ index of LG modes is contained in the waist of the beam or the curvature of the wavefront ~\cite{Bouchard2018}. Another factor contributing to the reduced visibility is slight misalignment in the system. We see that the average detection efficiency is approximately $40$\% and is nearly mode-independent. We can still observe a small drop in efficiency for higher values of the $p$ index, which is also attributed to the dependence of the beam waist on the orthogonality of the $p$ modes. However, we can see that for a fixed value $p$, the detection efficiency for all the ${\ell}$ values remain constant. If we remove the inherent losses in the system, namely from the SLM due to the additional blaze grating to remove the interference from the undiffracted light on the three planes ($\sim 80$\% in each plane) and the single-mode fiber coupling efficiency ($>$ $80$\%), we can see that the detection system is near perfect and lossless ~\cite{Hiekkamaki2019}.

\section{Spatial decomposition of photon pairs in different frequency bands, beam parameters of pump photon field and crystal lengths}\label{section:supp_NB}
\begin{figure}[htbp]
\includegraphics[scale=0.82]{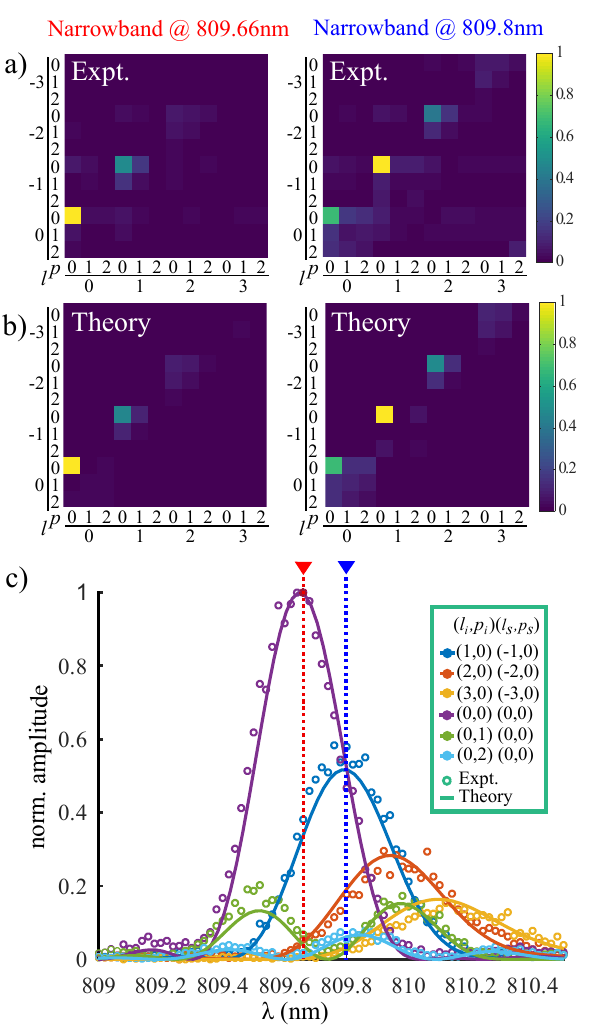}
\caption{Narrow band spatial decomposition around different central wavelengths: a) Experimental results around $\lambda=809.66$\,nm (red) and $\lambda=809.8$\,nm (blue). (b) Theoretical predictions for the narrow band spatial decomposition around $\lambda=809.66$\,nm (red) and $\lambda=809.8$\,nm (blue). (c) Joint spectrum of different joint spatial modes where the spectral filtering, around $\lambda=809.66$\,nm is shown with the red arrow and around $\lambda=809.8$\,nm is shown with the blue arrow. (Colors in online version).
} 
\label{fig:NB_diff_wavelengths}
\end{figure}

We have seen the effect of narrowband spectral filtering on spatial decomposition in the main section, where the filtering was performed around the center wavelength of the spectrum of the photons occupying the fundamental Gaussian mode. As a result, we could suppress the contribution of several higher-order modes as the central wavelengths of these modes, which carry higher transverse momenta, were shifted to higher wavelengths. With our experimental realizations closely matching the theory, we performed the spatial decomposition at a different frequency band slightly away from the central wavelength. The experimental setup is the same as given in the main text, with the only difference that the monochromator is set to measure around $\lambda=809.8$ $nm$, while it was $\lambda=809.66$ $nm$ in the main section. The narrow-band spatial decomposition obtained around $\lambda=809.8$ $nm$ compared to that around $\lambda=809.66$ $nm$ along with the expected results from the theory is shown in Fig. ~\ref{fig:NB_diff_wavelengths}. We can clearly see that the experimental results agree well with the theoretical predictions. While looking into the joint spatial amplitudes for different modes, we can see that for the spatial decomposition around $\lambda=809.8$ $nm$, the fundamental is no longer the brightest mode and here the combination of $l_s=1,p_s=0$ and $l_i=-1,p_i=0$ or vice versa has the brightest joint spatial amplitude. The spectral filtering at a different wavelength here suppresses the contribution of modes with lower transverse momenta and that is why the higher order modes can be brighter than the fundamental Gaussian mode. This is visible from the joint spectrum of the different spatial modes (Fig. ~\ref{fig:NB_diff_wavelengths}.c). 

\begin{figure}[!t]
\includegraphics[scale=0.82]{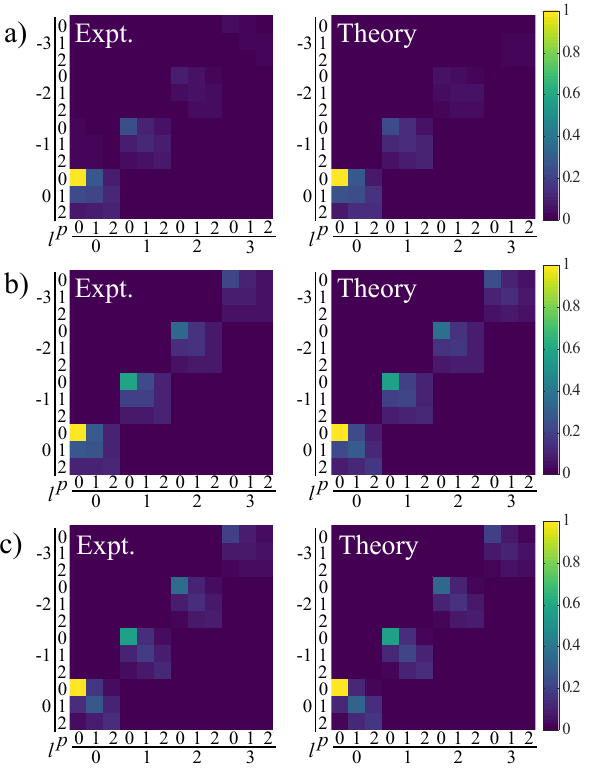}
\caption{Broadband spatial decomposition for different crystal lengths and pump beam waists: a) L= 20 mm,  $w_p\approx20$\,µm, b) L= 20 mm,  $w_p\approx60$\,µm, c) L= 10 mm,  $w_p\approx45$\,µm. The collection waist for the photon fields is set to $w_s\approx30$\,µm in all the three cases. (Colors in online version).
} 
\label{fig:spd_wps}
\end{figure}

Now we test the predictions of our theoretical model for the broadband spatial decomposition of photon pairs at different beam parameters for the pump photon field and also for two different crystal lengths. In the main text, we looked at the broadband spatial decomposition for a PPKTP crystal of length L= 20 mm, pump beam waist $w_p\approx60$\,µm and collection waist $w_s\approx30$\,µm. In Fig. ~\ref{fig:spd_wps}.a and Fig. ~\ref{fig:spd_wps}.b, we compare the experimental results and theoretical predictions for a tight and loose pump focusing, here $w_p\approx20$\,µm and $w_p\approx60$\,µm respectively. 

We can see that the relative weights for different joint spatial mode detections obtained in the experiment agree well with the theory. For a smaller beam waist for the pump, the contributions of the higher OAM modes are very less than those of the Gaussian mode. 
These parameters are important in applications where spatial entanglement has to be minimized so that entanglement in different degrees of freedom can be enhanced.  Fig. ~\ref{fig:spd_wps}.c shows the broadband spatial decomposition for the crystal length L= 10 mm, pump beam waist $w_p\approx45$\,µm and collection waist $w_s\approx30$\,µm. Here, we used a crystal that is different from that in the previous cases, and the experimental results are in accordance with the theory. Another interesting feature here is with respect to the spiral bandwidth of the OAM modes. The weights of the anticorrelated OAM modes are same as in Fig. ~\ref{fig:spd_wps}.b. 

\section{Optimization of the OAM collection modes for spectral decoupling}\label{section:supp_optimization}
 
As mentioned in the main text, a general biphoton OAM state can be written as $\ket{u_\ell}\ket{v_{-\ell}}=\sum_{p_\mathrm{i},p_\mathrm{s}} A_{p_\mathrm{i}}B_{p_\mathrm{s}} \ket{p_\mathrm{i},\ell} \ket{p_\mathrm{s},-\ell}$. And the spectral amplitude can be calculated through $C_{u,v}^\ell(\lambda)=\sum_{p_\mathrm{i},p_\mathrm{s}}A_{p_\mathrm{i}}B_{p_\mathrm{s}} C^\ell_{p_\mathrm{i},p_\mathrm{s}}(\lambda)$.

This approach allows us to control the spectral correlations by selecting the appropriate values of the coefficients $A_{p_\mathrm{i}}$ and $B_{p_\mathrm{s}}$. We can choose a particular normalized target spectrum $\Phi_t(\lambda)$ ($\int d\lambda |\Phi_t(\lambda)|^2=1$) and optimize the coefficients $A_{p_i}$ and $B_{p_s}$ minimizing the cost function:
\begin{equation}\label{eq:spec_overlap}
 F= 1 - |\int d\lambda \Phi^*_t(\lambda)C_{u,v}^\ell(\lambda)|^2,   
\end{equation}
where $C_{u,v}^\ell(\lambda)$ is properly normalized in each iteration. We can choose to optimize the brightness of a particular state by minimizing: 

\begin{equation}\label{eq:F_bright}
    F_{\mathrm{bright}}=1-\frac{\int |C_{u,v}^\ell(\lambda)|^2d\lambda}{\sum_{p_i,p_s}\int |C_{p_i,p_s}^\ell(\lambda)|^2d\lambda}.
\end{equation}

Similarly to Eq.~\ref{eq:spec_overlap}, we can optimize the spectral overlap and equalize the probability amplitude of the different OAM states $\ell$ and $\ell'$ using:
\begin{equation}\label{eq:spec_overlap_2} 
    F_\mathrm{spect}=1-\frac{|\int (C_{u,v}^\ell)^*C_{u,v}^{\ell'}(\lambda)d\lambda|^2}{\int |C_{u,v}^\ell(\lambda)|^2 d\lambda \int |C_{u,v}^{\ell'}(\lambda)|^2 d\lambda}.
\end{equation}

\begin{figure*}[htbp]
\includegraphics[scale=0.75]{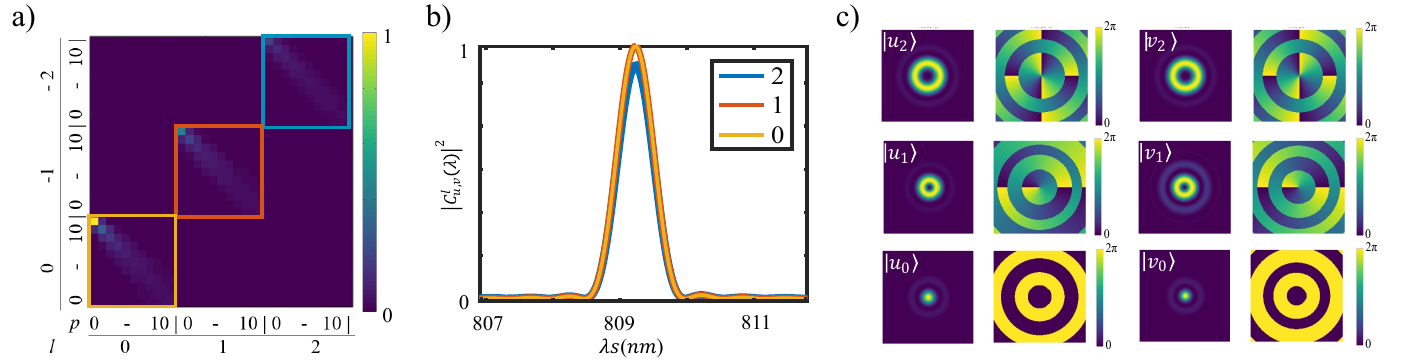}
\caption{a). Spatial decomposition in subspace $|\ell|=0,1,2$ and $p=0$ to $10$ using the pump and collection waists $w_p=50$\,µm and $w_s=50$\,µm. b) Spectrum of the calculated OAM modes that optimize the spectrum and amplitude overlap. c) Transverse spatial structure of the calculated OAM modes. (Colors in online version).} 
\label{fig:mode_optim}
\end{figure*}

For our simulation, we chose a $10$ $mm$ long type-II PPKTP crystal, and calculated the coefficients $C^\ell_{p_\mathrm{i},p_\mathrm{s}}(\lambda)$ using the pump and collection waists $w_\mathrm{p}=50$\,µm and $w_\mathrm{s}=50$\,µm. These are shown in Fig.~\ref{fig:mode_optim}.a for $|\ell|=0,1,2$ and $p=0$ to $10$. 
Now, we first use Eq.~\ref{eq:F_bright} to optimize the brightness for $|\ell|=2$ (subspace in the blue box). This can be done easily using the \textit{fminsearch} function of \textit{MATLAB}. We then use Eq.~\ref{eq:spec_overlap_2} for $\ell=2$ and $\ell'=0,1$ (subpaces in the yellow and orange boxes, respectively). The spectra of the calculated modes are plotted in Fig.~\ref{fig:mode_optim}.b showing almost perfect overlap and amplitude, potentially leading to a highly pure and bright 5-dimensional entangled state. Fig~\ref{fig:mode_optim}.c shows the reconstructed OAM modes $\ket{u_\ell}$ and $\ket{v_{-\ell}}$ in amplitude and phase. Interestingly, the optimal modes are not identical for signal and idler photons. This is especially noticeable for $|\ell|=1$, where the outer ring is more predominant. This can be further generalized by optimizing the pump mode, which can also be achieved by using our expression.

\bigskip
\normalem
\bibliographystyle{apsrev4-2}  
\bibliography{bibliography}


\end{document}